\documentclass{aa}  
\usepackage{natbib}
\usepackage{graphicx}
\graphicspath{ {images/} }
\usepackage{txfonts}
\usepackage{multirow}
\usepackage{hyperref}
\usepackage{color}

\begin{document}

   \title{Modeling optical and UV polarization of AGNs}

   \subtitle{IV. Polarization timing}

   \author{P. A. Rojas Lobos\inst{1},
           R. W. Goosmann\inst{1}
           F. Marin\inst{1}
           \and
           D. Savi{\'c}\inst{1,2}
          }

   \institute{1 Observatoire Astronomique de Strasbourg, Universit\'e de Strasbourg,
              11 rue de l\'{}Universit\'e, 67000 Strasbourg \\
              2 Astronomical Observatory of Belgrade, Volgina 7, 11060 Belgrade, Serbia\\
              \email{parltest@gmail.com}
             }

   \date{Received 08/06/2017; accepted 21/11/2017}

\abstract
% context (optional), leave it empty if necessary 
{Optical observations cannot resolve the structure of active galactic nuclei (AGN), and a unified model for AGN was inferred mostly from indirect methods, such as spectroscopy and variability studies. Optical reverberation mapping allowed us to constrain the spatial dimension of the broad emission line region and thereby to measure the mass of supermassive black holes. Recently, reverberation was also applied to the polarized signal emerging from different AGN components. In principle, this should allow us to measure the spatial dimensions of the sub-parsec reprocessing media.}
{We conduct numerical modeling of polarization reverberation and provide theoretical predictions for the polarization time lag induced by different AGN components. The model parameters are adjusted to the observational appearance of the Seyfert 1 galaxy NGC~4151.}
{We modeled scattering-induced polarization and tested different geometries for the circumnuclear dust component. Our tests included the effects of clumpiness and different dust prescriptions. To further extend the model, we also explored the effects of additional ionized winds stretched along the polar direction, and
of an equatorial scattering ring that is responsible for the polarization angle observed in pole-on AGN. The simulations were run using a time-dependent version of the {\sc stokes} code.}
{Our modeling confirms the previously found polarization characteristics as a function of the observer`s viewing angle. When the dust adopts a flared-disk geometry, the lags reveal a clear difference between type 1 and type 2 AGN. This distinction is less clear for a torus geometry where the time lag is more sensitive to the geometry and optical depth of the inner surface layers of
the funnel. The presence of a scattering equatorial ring and ionized outflows increased the recorded polarization time lags, and the polar outflows smooths out dependence on viewing angle, especially for the higher optical depth of the wind ($\tau$=0.3).}
{Together with other AGN observables, the polarization time lag places new, independent ``seismological'' constraints on the inner geometry of AGN. If we conduct time-dependent spectropolarimetric observing campaigns of AGN, this method has a high potential for a census of supermassive black holes.}


\maketitle
%
%________________________________________________________________
%

\section{Introduction}
\label{sec:intro}

Active galactic nuclei (AGN) are very compact and high-luminosity engines located at the center of their host galaxies. Since the innermost structures of AGN cannot be resolved by current UV,
optical, or IR observational techniques, morphological and kinematic models have to be derived indirectly, most often from broadband spectroscopy and variability studies.

In so-called thermal radio-quiet AGN, the emission is due to accretion onto a supermassive black hole (SMBH),with typical masses ${\rm M}_{\rm{SMBH}}$>${10^6 \rm M}_\odot$  \citep{1964ApJ...140..796S,1969Natur.223..690L}. The accretion disk produces the optical and UV continuum radiation \citep [see][]{1978Natur.272..706S, 1973A&A....24..337S, 1973A&A....29..179P}. The disk axis points out a preferred direction that for the remainder of this paper we call the polar direction of the system, to be distinguished from the equatorial directions that enclose the plane of the accretion disk. Observationally, thermal AGN can then be divided into two classes: type 1 sources are assumed
to be viewed at a polar line of sight and show Doppler-broadened Balmer emission lines produced in the so-called broad line region (BLR), whereas type 2 sources are viewed at equatorial lines of sight, at which the optical view of the BLR is obscured by a dusty region commonly called the dusty torus \citep{1993ARA&A..31..473A}. It extends out to a spatial scale of a few parsecs for a $10^7 {\rm M}_\odot$ SMBH. Only in very few objects do the outer parts of the circumnuclear dust start to be resolvable by mid-IR interferometry techniques \citep{2007A&A...474..837T, 2014A&A...563A..82T, 2009MNRAS.394.1325R}. 

\Citet{2009NewAR..53..140G} introduced the so-called bird's nest paradigm for the central region of thermal AGN, adopting the idea that its overall structure is clumpy. Clumpy models successfully reproduce the dust emission seen in the IR \citep{2008ApJ...685..160N,2015A&A...583A.120S,2010A&A...523A..27H}. The BLR is expected to be fragmented and turbulent, but radially stratified, as can be inferred from the velocity structure of low- and high-ionization emission lines \citep{1982ApJ...263...79G,2008MmSAI..79.1090G,2013ApJ...769...30G}. The BLR is expected to form inside the sublimation radius, possibly from inflowing torus material. It is therefore embedded in the torus funnel. Since the outer parts of the accretion disk around the SMBH must be self-gravitating, it is straightforward to identify it with the inner parts of the BLR. A coherent picture for the circumnuclear matter of AGN can be found with a morphology that is non-spherical, but symmetric with respect to the polar axis, while being fragmented on small spatial scales.

\begin{figure}[h]
  \centering
  \includegraphics[width=0.49\textwidth]{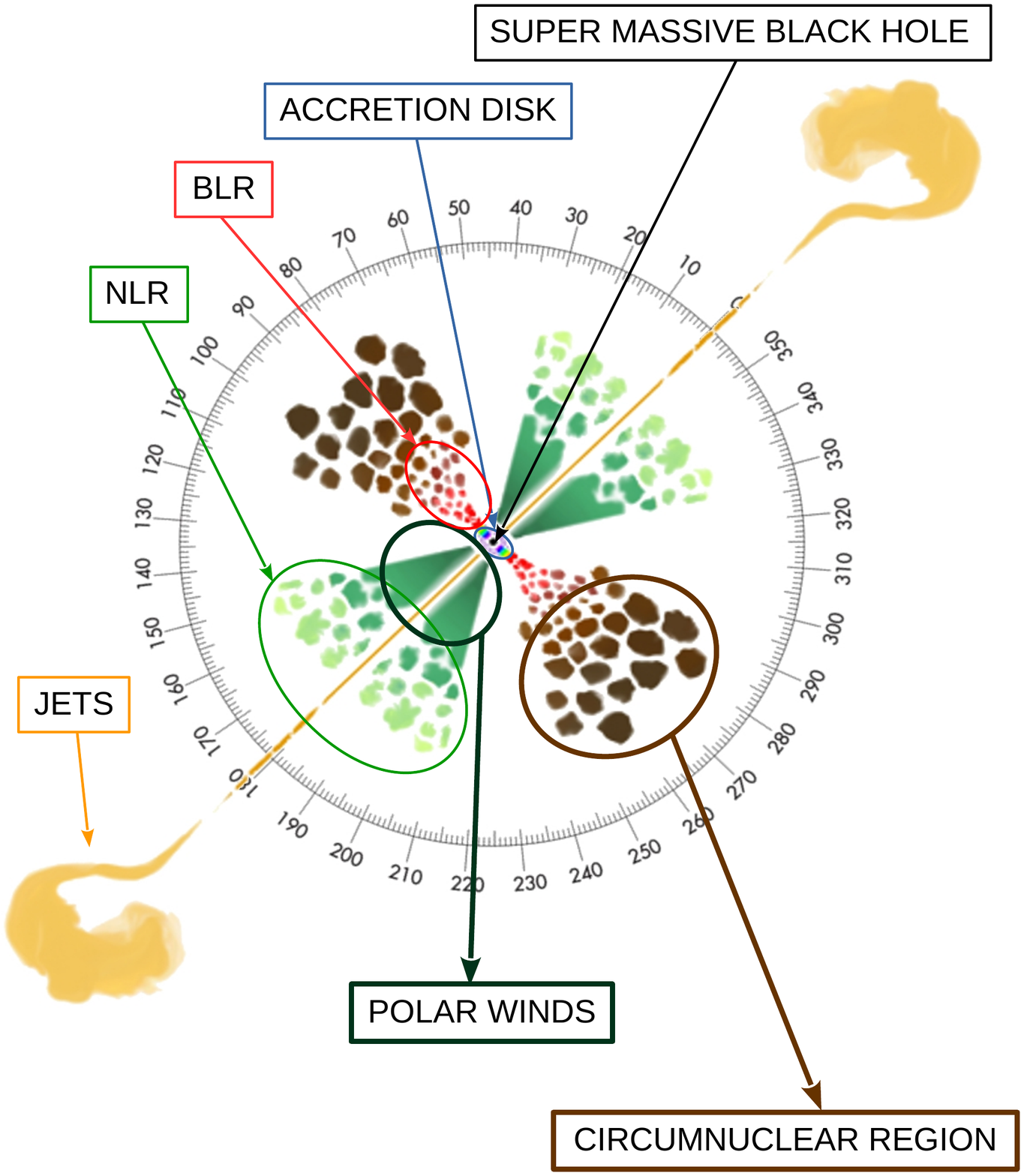}
  \caption{Unscaled sketch of the unified model of AGN, taken from \citet{2016MNRAS.460.3679M}. The cartoon shows that AGN with narrow and broad Balmer emission lines (type 1, $i = 0-60^\circ$) and those with only narrow lines (type 2, $i = 60-90^\circ$) derive from the same morphology, but are viewed from different angles with respect to the AGN axis. This is the original basis of the unified model. \label{fig:sketch}}.
\end{figure}

Not all matter that is pulled in by the black hole is finally accreted. Depending on the accretion state of a given thermal radio-quiet AGN, a significant fraction of matter may be ejected along the polar directions as moderate to ultrafast winds \citep{2006MNRAS.372.1275P, 2003MNRAS.345..657K, 2012MNRAS.422L...1T}. Depending on the distance from the SMBH, these winds are mildly to strongly ionized and scatter much radiation that is emitted by the central black hole. The common picture of the inner structure of AGN was recently described by \citet{2016MNRAS.460.3679M}, and the reproduction in Fig~\ref{fig:sketch} depicts the regions inside the AGN (not to scale): the black hole sits at the center and is surrounded by the accretion disk, the BLR, and the circumnuclear dust. Along the polar directions, ionized winds and the narrow line region (NLR) are located, which in radio-loud objects may be pierced by collimated jets.

Polarimetry is a valuable tool to probe the innermost structures of AGN. In the past, optical spectropolarimetry gave proof of the unified theory of AGN when \citet{1985ApJ...297..621A} discovered polarized broad Balmer lines in Seyfert 2 galaxies and thus pointed out a relation between the observer's inclination with respect to the polar axis and the position angle of optical continuum polarization in type 1 and type 2 AGN. It was shown in that the typical polarization properties in type 1 objects can be produced by scattering in a flattened distribution of material situated around the accretion disk \citep{1984ApJ...278..499A,2004MNRAS.350..140S,2007A&A...465..129G}.

More recently, AGN polarimetry has been brought to a different level by introducing the technique of polarization reverberation mapping (\citet{{2012ApJ...749..148G}} for the Seyfert
1 nucleus NGC~4151). In this method, the time-dependent total (polarized+unpolarized) continuum emission is cross-correlated with the polarized emission of a given AGN. A characteristic time delay is found between the two components. When we assume that the continuum polarization is induced by scattering of the continuum radiation by surrounding structures, the reverberation places constraints on the light travel distance between the emitting source and the reprocessing mirror(s).

In this work, we present initial modeling to compare with current and future campaigns of polarization reverberation mapping. We conduct radiative transfer simulations for the continuum radiation in different AGN geometries, taking the time dependence of the polarized radiation in the optical band into account. The goal of this research is to numerically explore the new method and
indirectly resolve the innermost region in AGN by polarization reverberation. To do this, we provide simulated results for polarization time lags from different AGN constituents as a function of inclination. The paper is organized as follows: in Sect.~\ref{sec:model} we present the details of our model and specific parametrization. In Sect.~\ref{sec:results} we present and analyze our results. Finally, in Sect.~\ref{sec:discuss} we discuss our results and follow-up projects.

\begin{table*}
        \caption{Model setup, baseline. Parameters of three scattering regions 
        with uniform and clumpy dust distribution for the Mie scattering 
        and free electron distribution for Thomson scattering}
        % title of Table
        \label{tab:param}      % is used to refer this table in the text
        \centering                          % used for centering table
        \begin{tabular}{  p{5cm}  p{1.5cm}  p{2.9cm}  p{3.2cm}  p{2.9cm} }     
        \hline%\hline                 % inserts double horizontal lines

        MODELS & TORUS & FLARED DISK & SCATTERING RING & POLAR WIND \\    % table heading 
        \hline                        % inserts single horizontal line
        DUSTY & & & & \\
        Inner radius (pc) & 0.061  & 0.061 & 0.0067 & 0.0067 \\
        Outer radius (pc) &15.061 &15.061 & 0.0577 & 30.0067 \\ 
        Half opening angle & 60$^{\circ}$ & 60$^{\circ}$ & 60$^{\circ}$ & 30$^{\circ}$ \\         
        Equatorial optical depth & 150 & 150 & 1 & 0.03 / 0.3 \\
        Scattering & Mie  & Mie  & Thomson & Thomson  \\
        \hline
        CLUMPY & & & & \\
        Inner radius (pc) & 0.061 & 0.061 & &   \\
        Outer radius (pc) & 15.061 & 15.061 & &   \\ 
        Half opening angle & 60$^{\circ}$ & 60$^{\circ}$ & &  \\     
        Optical depth per cloud & 50 & 50 &  & \\
        Clouds radius (pc) & 0.2 & 0.65 &  & \\
        Filling factor & 25\% & 25\% &  & \\
        Scattering & Mie  & Mie  &   & \\ 
        \hline%\hline   

        \hline                                   %inserts single line
        \end{tabular}
\end{table*}

\section{Modeling the circumnuclear dust scattering}
\label{sec:model}
\begin{figure}[h]
        \centering
        \includegraphics[width=0.49\textwidth]{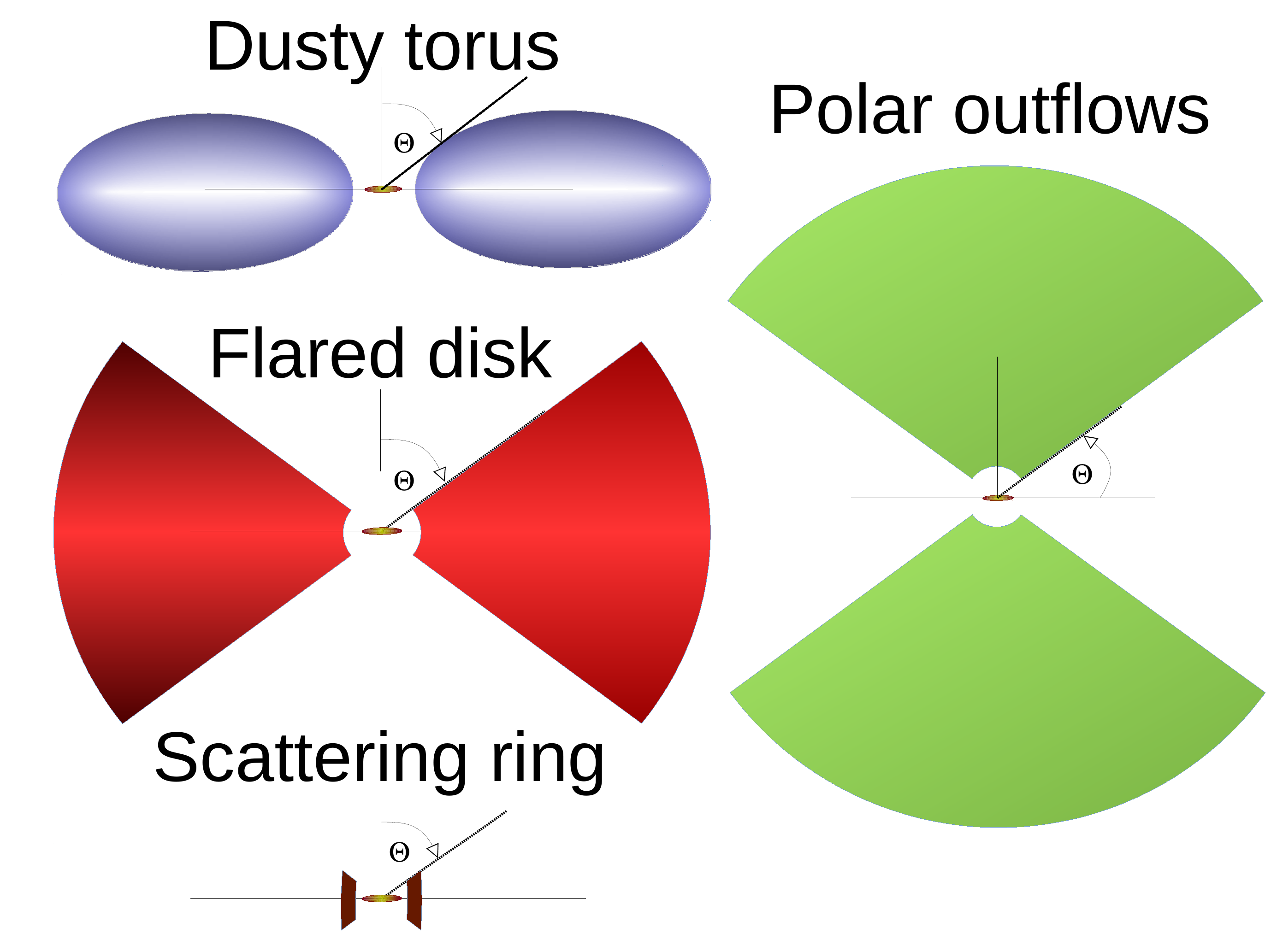}    
        \caption{Geometry scheme of the models. \textit{Left top}: classical torus geometry (blue) with elliptical cross section. \textit{Left middle}: the extended flared-disk geometry (red) with the wedge-shaped  cross section. \textit{Left bottom}: Scattering ring (brown). \textit{Right}: polar outflows (green).\label{fig:sketch_UM}}
\end{figure}

Following the observational example of \citet{2012ApJ...749..148G}, we here attempt to theoretically investigate the details of the inner AGN structure by studying the expected time lag between the total and the polarized continuum flux. The presence of the obscuring torus on the line of sight should be derived from the time lag, together with its geometrical shape and composition. Figure~\ref{fig:sketch_UM} shows a sketch of our different modeling options. In the following, we call ``dusty-torus'' a toroidal region with an elliptical cross section, and ``flared-disk'' a region with a wedge-shaped cross section. We investigate polarization timing for four versions of a circumnuclear geometry: a uniform-density dusty torus, a uniform-density flared disk, a clumpy torus, and a clumpy flared disk. Every geometry is tested for two types of optically thick dust: AGN-dust \citep{2004ApJ...616..147G} and Milky Way dust \citep[so-called MRN dust, according to][]{1977ApJ...217..425M} at different optical depths.
We increase the complexity of the model by adding ionized outflows in the polar direction in Sect.~\ref{Polar_outflows} and a scattering ring, responsible for the specific polarization position angle seen in type 1 AGN \citep{2004MNRAS.350..140S}, in Sect.~\ref{Ring}. The latter region is fully ionized and is located between the source and the circumnuclear dust region. As the model setup becomes more sophisticated, we follow the
evolution of the polarization time lags.

We applied an extended version of the Monte Carlo radiative transfer code {\sc stokes}. This version includes the timing information of the registered photons. The code was initially written by \citet{2007A&A...465..129G} and further developed by \citet{2012A&A...548A.121M} and \citet{2015A&A...577A..66M}; an intermediate version of the code is publicly available \footnote{http://stokes-program.info/}. It simulates radiative transfer in different geometries for emitting and scattering structures and determines the net Stokes parameters as a function of wavelength, time, and the observer's line of
sight. Subsequently, it is possible to obtain the polarization fraction, position angle, polarized flux, and total flux from the Stokes parameters.
For compatibility with paper I \citep{2007A&A...465..129G}, we define the polarization position angle to be 0$^{\circ}$ when the polarization angle is perpendicular to the projected symmetry axis of the model. When the polarization angle is equal to 90$^{\circ}$, it is said to be parallel to the projected symmetry axis of the model. To compute the time lags, the code compares the path between the direct flux from the source and the trajectory of the photons that have been scattered from the dust and/or electron regions. For each viewing direction, the time lag is measured relative to the plane of sky, which goes through the origin of the central source and is oriented perpendicularly to the line
of sight. The lag is expressed in $\rm{pc/[c]}$ and represents the light traveling distance.

We modeled the flared-disk and toroidal geometries shown in Fig~\ref{fig:sketch_UM} assuming uniform-density and clumpy dust distributions. Two types of dust were tested: AGN-dust composed of 85\% of silicate and 15\% of graphite, and Milky Way dust composed of 62.5\% of silicate and 37.5\% of graphite. The grain radii lay between 0.005$\mu$ (0.005$\mu$) and 0.25$\mu$ (0.2$\mu$), with a distribution $n(a)~\propto~a^{s}$ with $s= -3.5$ ($s = -2.05$) for Milky Way (AGN) dust. The filling factor of the clumpy regions was set to 25\% \citep{2015A&A...577A..66M}. At the center we defined a point source with unpolarized flux that emitted a power law spectrum $F_{\rm\nu}\propto\nu^{-\alpha}$ with ${\alpha}=1$. For the ionized regions we assumed an inner radius of 8~light days such as found by \citet{2012ApJ...749..148G} in the case of NGC~4151, but without detailed modeling.
In the case of the circumnuclear dusty regions, we adopted the observational constraints derived by near-IR observation of NGC~4151, namely 0.061 pc, according to \citet{2013ApJ...775L..36K}. 

The outer radius of the torus and its half opening angle were based on the findings by \citet{2003MNRAS.340..733R}, Table 2, who conducted near-IR polarimetry observations of the same object. We defined an external radius of 15.061~pc and a half opening angle of 60$^{\circ}$ from the disk symmetry axis. The dust is opaque, with a radial optical depth of $\sim 150$ inside the equatorial plane. The modeling results can be evaluated over the optical wavelength range and as a function of the observer's viewing angle.
        
In a second step of this study, we added polar winds to this baseline model. They have the geometrical shape of double-cones and are assumed to be ionized. In the model they are filled with electrons with a radial Thomson optical depth of 0.03 and 0.3. 
The winds have a common interface with the circumnuclear dust at the inclination of its horizon. Finally, we added a third region, an equatorial scattering ring located between the source and the circumnuclear dusty region. This area, modeled as a flared disk, was found to be necessary to reproduce the parallel polarization position angle observed in the majority of Seyfert 1s, see \citet{2004MNRAS.350..140S}. Following the composition constraints from Paper~II, we opted for a radial optical depth of unity. The inner radius of the scattering ring was set to 0.0067 pc (i.e., eight lights days, \citealt{2012ApJ...749..148G}) and the outer radius to 0.0577 pc, so that the disk and the torus did not mix (see Table~\ref{tab:param} for a summary of the parameters).

\begin{figure}[h]
  \centering
  \includegraphics[width=0.50\textwidth, origin =c]{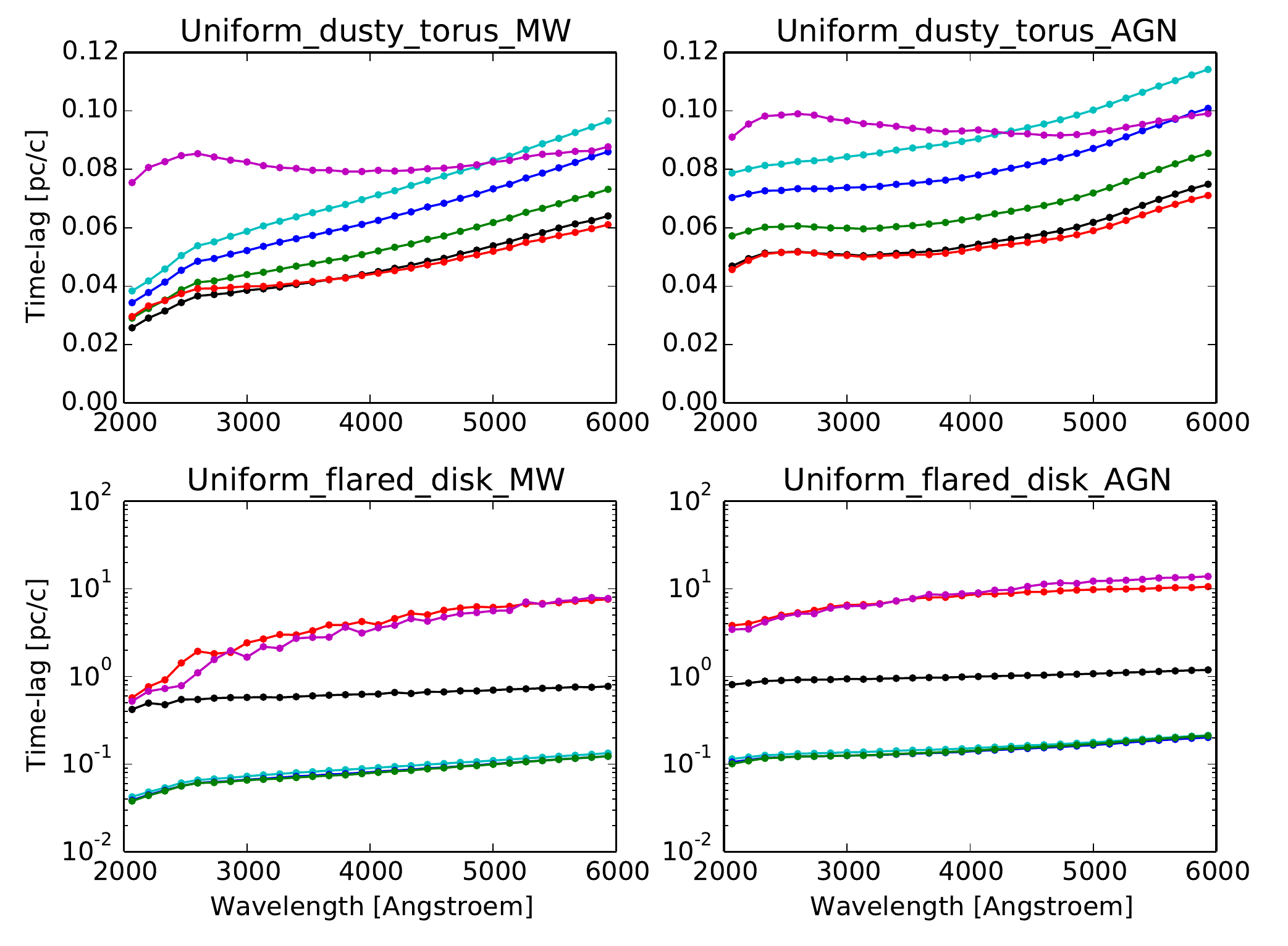}
  \caption{Comparison of the time lags induced by the Milky Way and by AGN dust prescription for the 
  case of a uniform dusty torus and a uniform flared disk. The plots show the time lag as a function of
    wavelength for different viewing angles: the magenta dots
    denote  $\theta$ = 79.9 $^{\circ}$, and the red dots $\theta$ = 71.0
    $^{\circ}$, which represents a type 2 view.  The black dots show
    $\theta$ = 61.64 $^{\circ}$, a view close to the torus horizon.
    Type 1 views are shown in green ($\theta$ = 51.3$^{\circ}$), blue
    ($\theta$ = 39.19$^{\circ}$),  and cyan($\theta$ =
    28.9$^{\circ}$).  \label{fig:TL-dust}}
\end{figure}    

\section{Results and analysis}
\label{sec:results}
        
All modeling cases were tested for the two different dust prescriptions: Milky Way dust \citep{1977ApJ...217..425M}, and AGN-dust \citep{2004ApJ...616..147G}. A  detailed comparison of the effect of the dust prescription on the  polarization results was given in \citet{2007A&A...465..129G}. The new simulated observable added in this work, the time lag of the polarized emission, is not much affected when the dust prescription is modified (Fig~\ref{fig:TL-dust}). The small differences between the two cases are due to the specific albedo and scattering phase functions of the two dust types. Since there is no major effect on the results, we only present modeling for the case of standard Milky Way dust in this paper.

\begin{figure}[h]
  \centering
  \includegraphics[width=0.49\textwidth, origin =c]{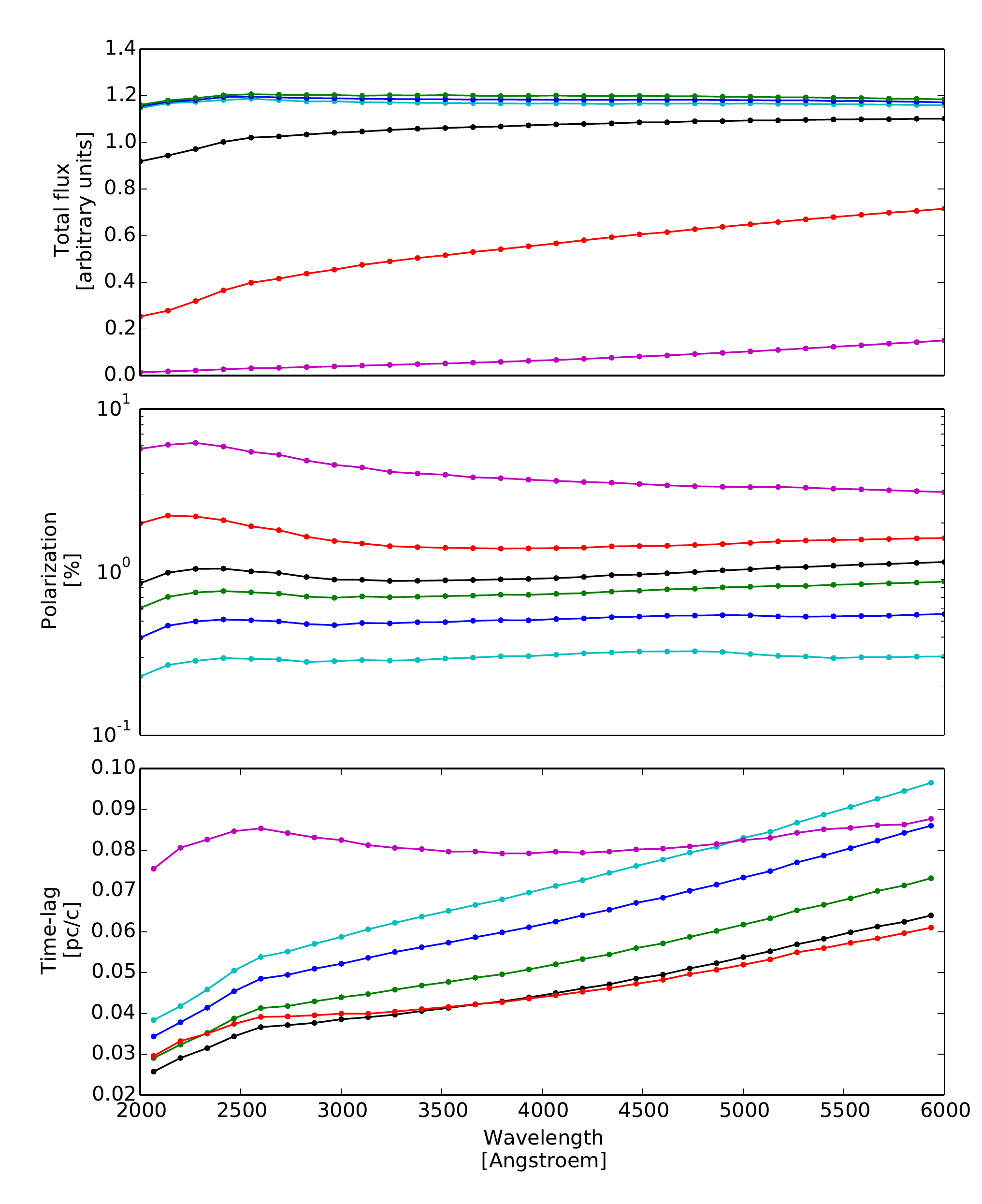} 
  \caption{Results of {\sc stokes} time-dependence version for
the     uniform dusty torus geometry. \textit{Top}: total
    flux, polarization percentage, and time lag as a function of
    wavelength and for different viewing angles: the magenta dots
    denote  $\theta$ = 79.9 $^{\circ}$, and the red dots $\theta$ = 71.0
    $^{\circ}$, which represents a type 2 view.  The black dots show
    $\theta$ = 61.64 $^{\circ}$, a view close to the torus horizon.
    Type 1 views are shown in green ($\theta$ = 51.3$^{\circ}$), blue
    ($\theta$ = 39.19$^{\circ}$),  and cyan($\theta$ =
    28.9$^{\circ}$). \textit{Middle}: percentage of
    polarization. \textit{Bottom}:
    time lag. \label{fig:TF-PO-TL-lambda}}
\end{figure}

The response in terms of flux, polarization, and time lag only mildly changes with wavelength (Fig.~\ref{fig:TF-PO-TL-lambda}). Only at type 2 viewing angles does the uniform dusty torus show a diminution in flux and an increase in polarization around the 2175\AA~ absorption feature. Overall, the results shown in Fig.~4 confirm our previous work: for a type 1 inclination, we obtain a higher flux level and lower polarization than at type 2 viewing angles. The new simulated observable in this work is the average time lag of the radiation. We note that this time lag is taken relative to the direct emission coming from the central source. The units are set to $pc/c$, which allows us to interpret the time lag in a straightforward manner in terms of light traveling distances. This time lag is always higher at type 2 viewing directions than for type 1, which is expected because below the torus horizon, we only detect (multiply) scattered radiation that has taken a non-direct path from the central source to the observer.

Except for these features, the results in terms of spectral flux, polarization, and time lag do not vary much with wavelength. For the remainder of this paper we therefore show our results for the V band centered on 5500\AA~and as a function of viewing angle. All models show significant differences between face-on and edge-on views.

\begin{figure}[h]
  \centering
  \includegraphics[width=0.49\textwidth, origin =c]{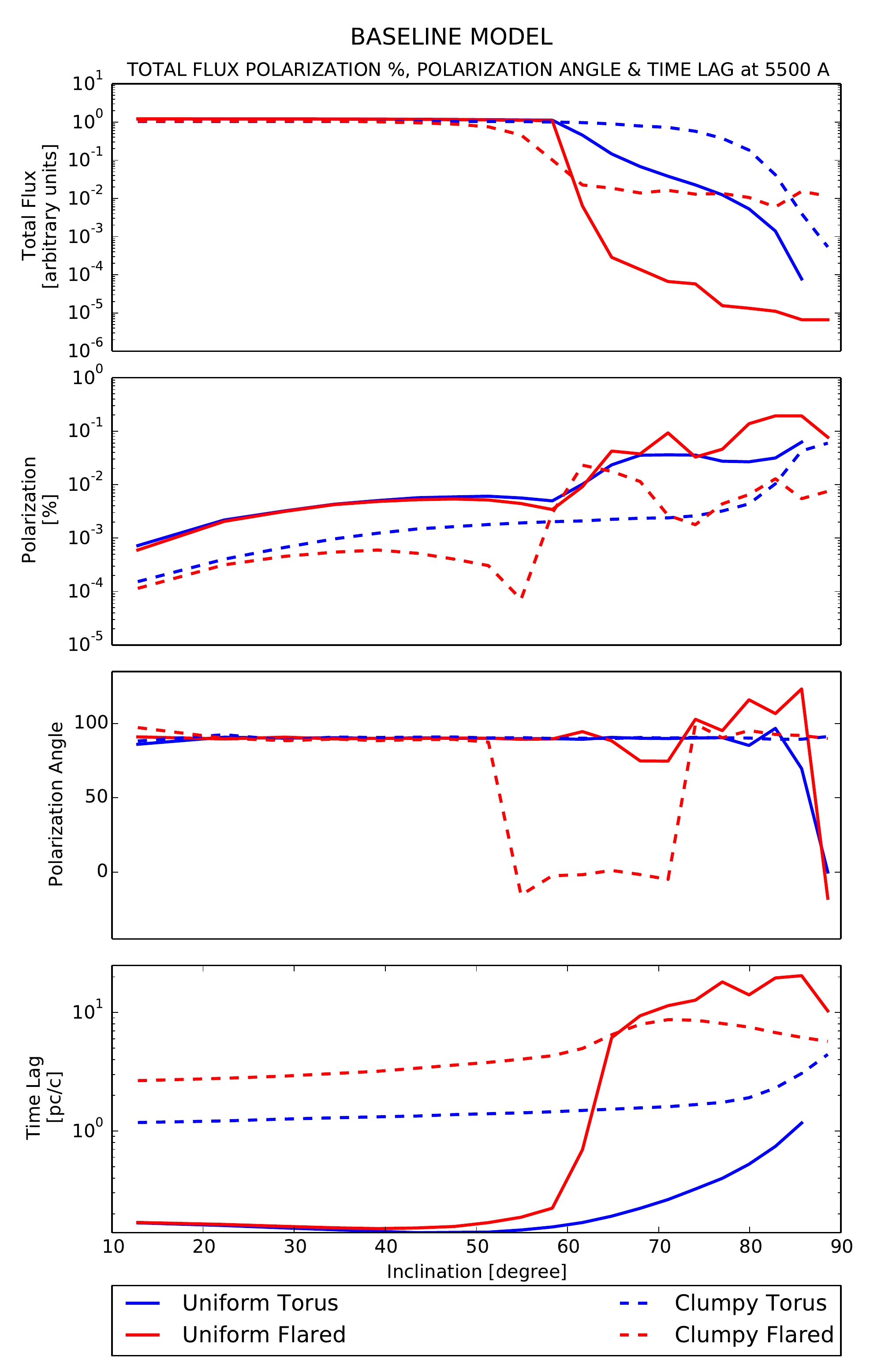} 
  \caption{Stokes results for V-band normalized flux,
    percentage of polarization, polarization angle, and time lag for
    different viewing angles. The blue line represents the uniform
    dusty torus model, and the blue dashed line the clumpy dusty torus,
    the red line the uniform flared disk, and the red dashed line
    represents the clumpy flared disk.\label{fig:no_winds}}
\end{figure}

\subsection{Baseline model for circumnuclear dust}

Our baseline model contains the axisymmetric dust region with a half opening angle of 60$^{\circ}$ as measured from the axis (see Fig.~\ref{fig:sketch_UM}). The modeling results are plotted in Fig.~\ref{fig:no_winds}. At face-on inclinations, all four models show roughly the same spectral shape in total flux, polarization degree, and time lag. The normalization of the spectral flux is almost the same for all four morphological setups because at these viewing angles the radiation predominantly comes from the central source. Still, a face-on observer finds a higher time lag and lower polarization degree for the clumpy dust distribution than for the uniform density cases. This is due to radiation scattered into the type 1 lines of sight. In a clumpy circumnuclear environment, the escaping photons experience on average a higher number of scatterings before they end up on a type 1 line of sight \citep[see][]{2015A&A...577A..66M}. For optically thick uniform-density regions, the scattered radiation at face-on viewing angles is mostly dominated by the first-scattering component that is more strongly polarized and accumulates less of a time lag.

At edge-on inclination, the observer's line of sight is obscured by dust, and the flux strongly decreases. The polarization degree slowly increases with increasing viewing angle for all four geometries. At all type 2 viewing angles, the time lag accumulated for a clumpy or uniform torus geometry is much shorter than for the case of a flared dust distribution.

This difference is related to the shape of the inner surfaces. For a torus, the inner surface is convex, while for the flared disk it is parametrized as a concave shape. As discussed in \citet{2007A&A...465..129G}, when comparing the obscuration efficiency between compact and extended dusty tori, we find again here that a more convex shape of the inner surfaces makes it more likely for a photon to escape after only a few scattering events. Therefore, the accumulated time lag for the case of a toroidal dust distribution remains short even when the distribution is clumpy.

For the case of our flared geometry, the inner surface is parametrized as a sphere segment against which the photons are injected from the inside. In general, multiple backscattering inside this sphere segment is necessary for the photons to finally escape, and if they do, the escape cone is strictly limited by the half opening angle of the flared disk as measured from its symmetry axis. Figure~5 shows that the rise in time lag and polarization degree is steep around this inclination for the flared dusty disk, while for the toroidal geometry, the rise is much more shallow. For clumpy dust distributions, the horizon of the torus and of the flared disk is more blurry, which decreases the slope of the polarization degree and the time lag even further. The polarization position angle remains at 90$^{\circ}$ (i.e., parallel to the projected symmetry axis, following the usual convention in this paper series) over all viewing angles for models without polar winds, with the exception of a clumpy flared geometry, for which the polarization position angle stays at 90$^{\circ}$ at face-on viewing angles, then shifts to 0$^\circ$ (perpendicular polarization) around the transition between type 1 and type 2 to return to parallel polarization at more extreme type 2 viewing angles. We argue that this behavior is related to the actual size of the clumps and to their distribution along the line of sight.

\begin{figure*}[ht!]
  \centering
  \includegraphics[width=0.49\textwidth, origin =c]{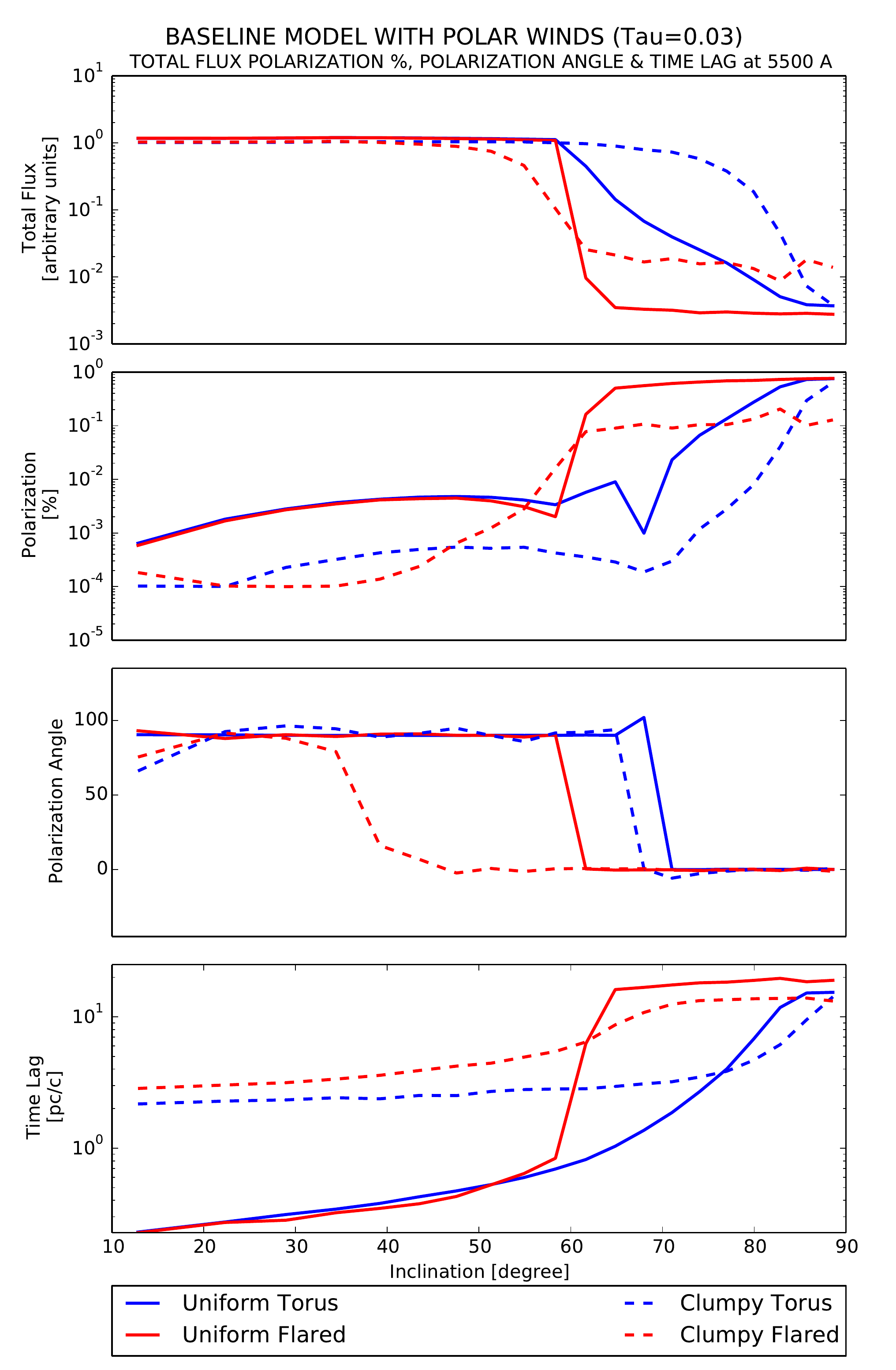}  
  \hfill
  \includegraphics[width=0.49\textwidth, origin =c]{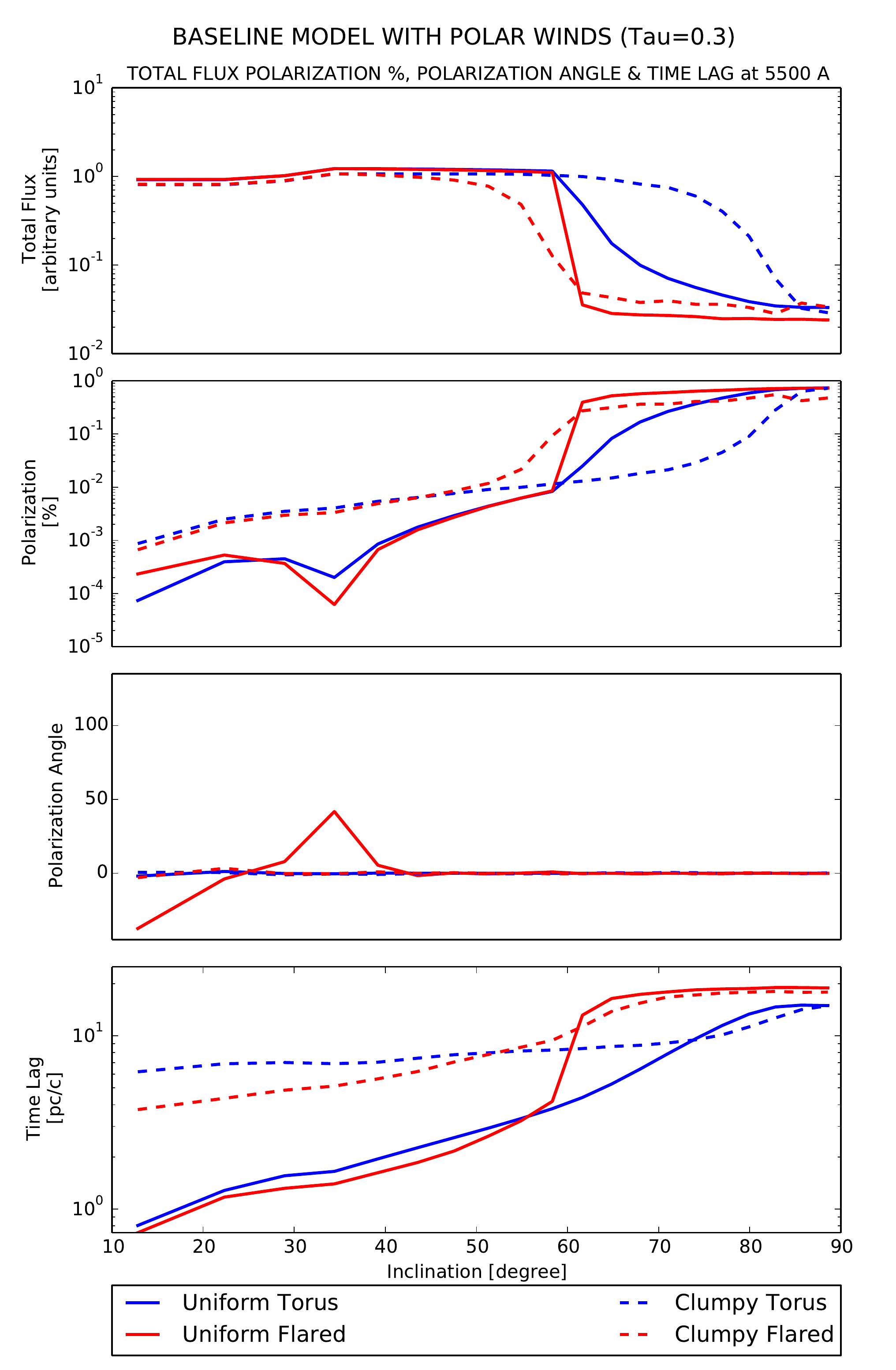}   
  \caption{Integrated and normalized flux, percentage of polarization,
    polarization angle, and  time lag for different viewing angles for baseline model, with an ionized polar wind of 60$^{\circ}$ half opening angle
    Thomson optical depth of $\tau =0.03$ (Left) and $\tau =0.3$ (right).}
  \label{fig:winds}
\end{figure*}

\begin{figure}[ht!]
  \centering
  \includegraphics[width=0.49\textwidth,origin =c]{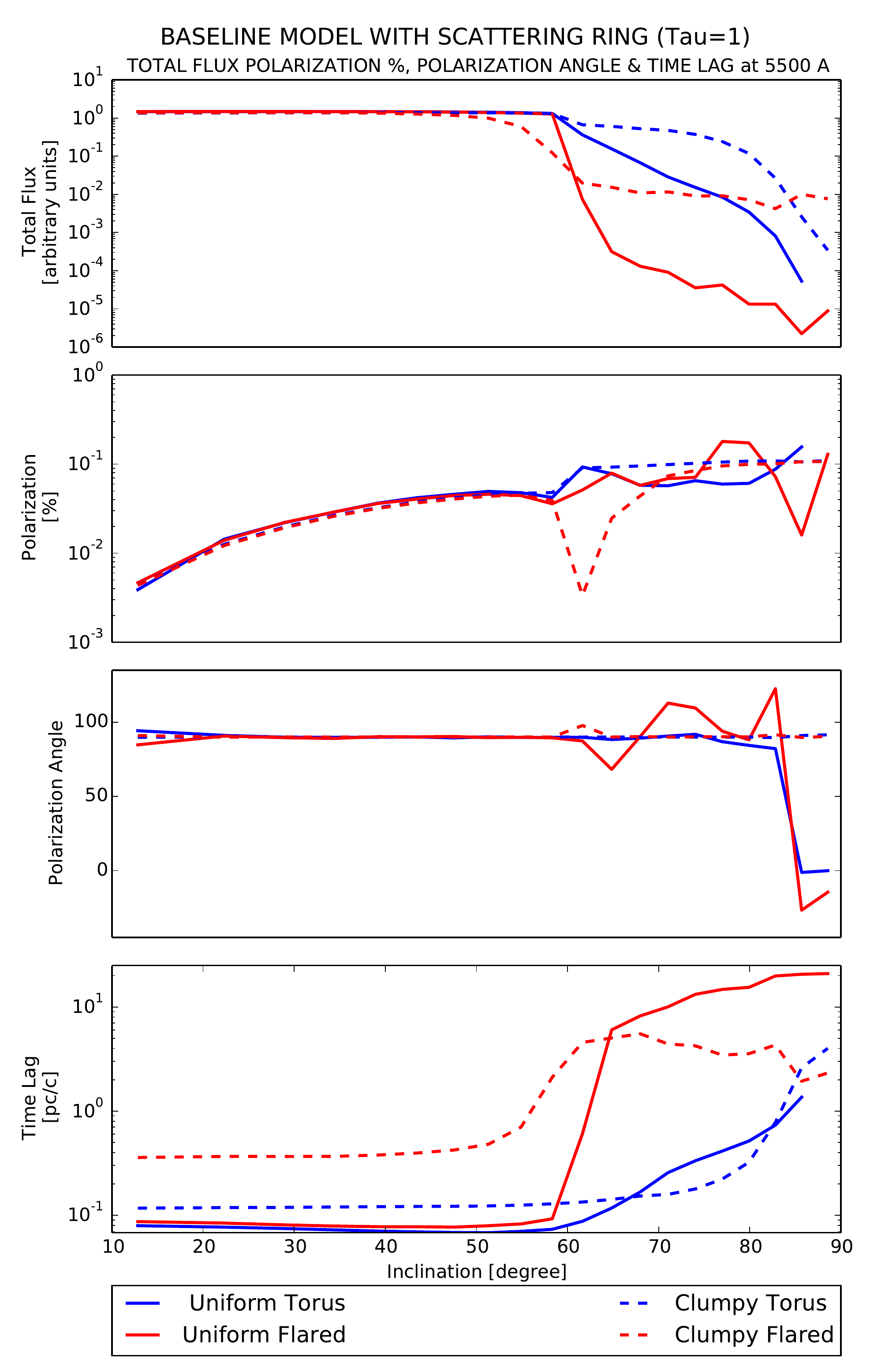} 
  \caption{Baseline model with equatorial ring. Stokes results for V-band normalized flux,
    percentage of polarization, polarization angle, and time lag for
    different viewing angles. The blue line represents the uniform
    dusty torus model, the blue dashed line the clumpy dusty torus,
    the red line the uniform flared disk, and the red dashed line
    represents the clumpy flared disk.\label{fig:baselineEQ}}
\end{figure}

\subsection{Adding polar outflows to the baseline model}
\label{Polar_outflows}
Next, we added ionized winds stretched along the symmetry axis to our model. As in previous work of this series, the wind geometry was realized by a double-cone filled with electrons of a radial optical depth of 0.03 or 0.3 (see Fig.~\ref{fig:winds}). When adding a low-density wind, the type 1 cases exhibit almost the same spectral flux level and polarization degree as for the baseline models. The additional scattering in the polar direction has a more significant effect on the time lag recorded at face-on viewing angles. It increases by a factor of roughly 2 when compared to the case without polar winds. When we add low-density polar winds, the polarization angle changes from 90$^{\circ}$ (parallel) at face-on view to 0$^{\circ}$ (perpendicular) in face-on for uniform models, while for a clumpy distribution, the polarization anlge remains at 0$^{\circ}$ for both geometries at all viewing angles.

\begin{figure*}[ht!]
  \centering
  \includegraphics[width=0.49\textwidth, origin =c]{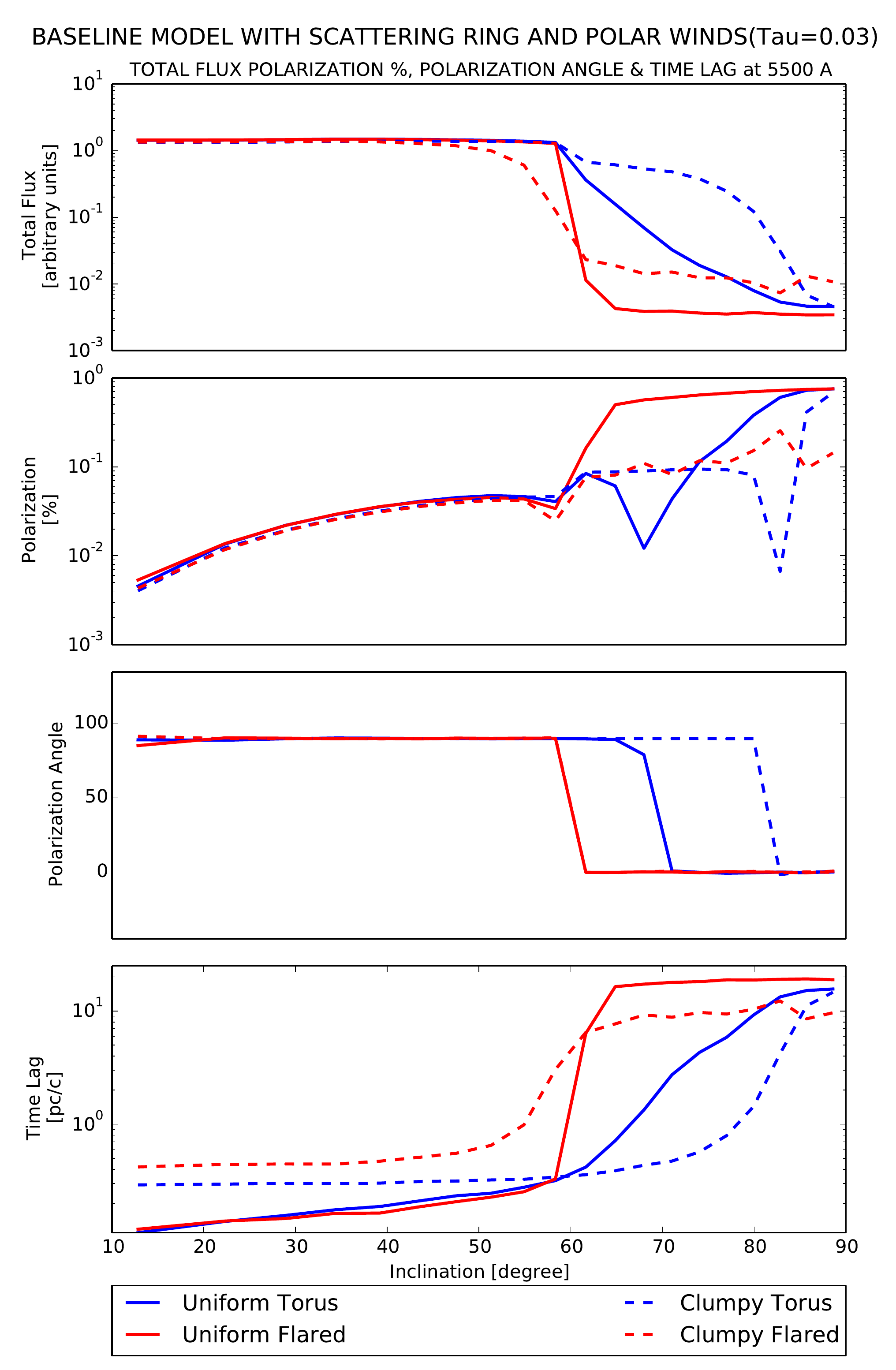}  
  \hfill
  \includegraphics[width=0.49\textwidth, origin =c]{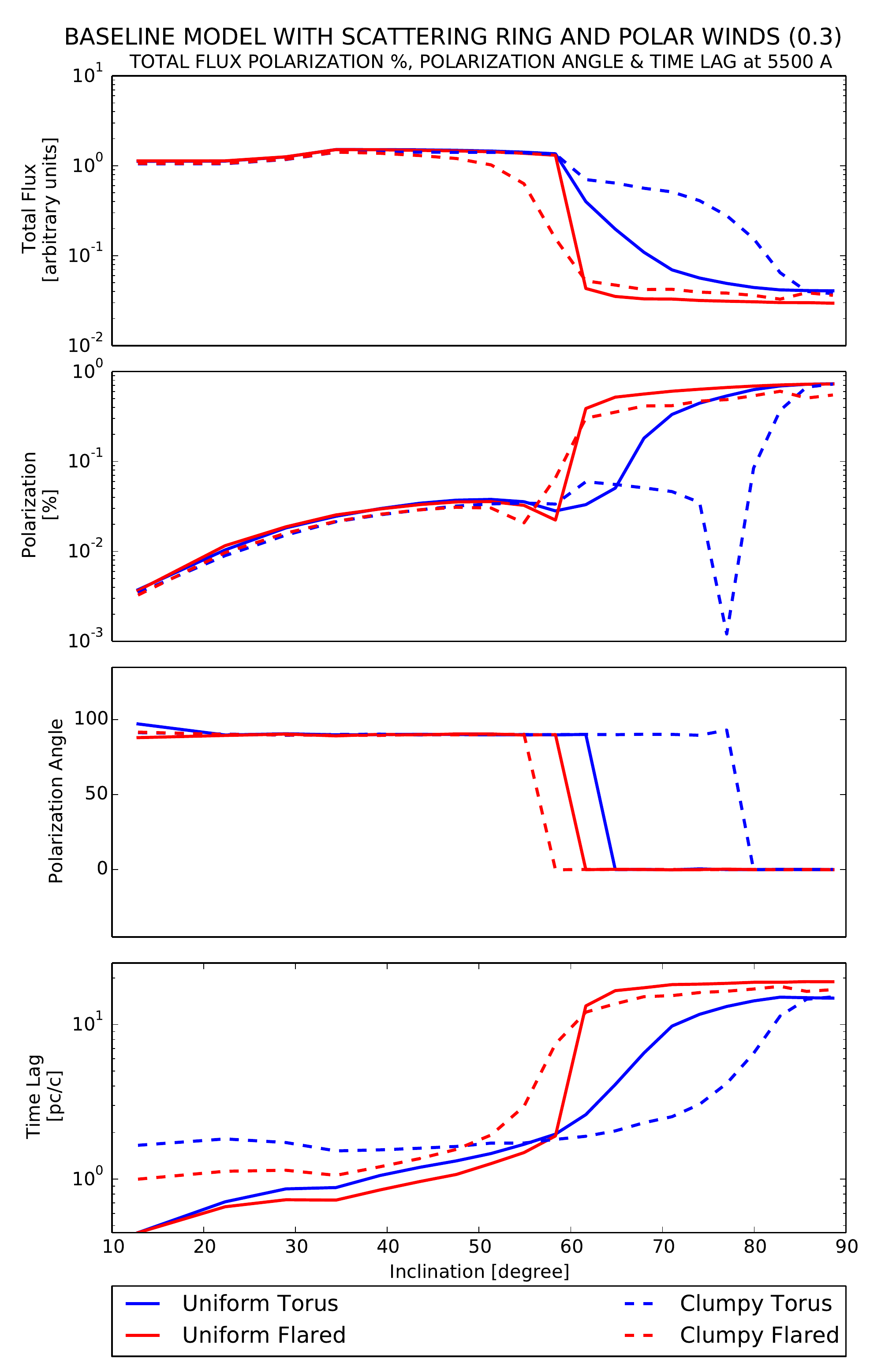}   
  \caption{Integrated and normalized flux, percentage of polarization,
    polarization angle, and  time lag for different viewing angles for the baseline model with a scattering ring, with an ionized polar wind of 60$^{\circ}$ half opening angle, a
    Thomson optical depth of $\tau =0.03$ (left), and $\tau =0.3$ (right).}
  \label{fig:windsEQ}
\end{figure*}

The situation is slightly more spectacular for the edge-on case. The presence of even a low-density wind scatters a significant part of the radiation ``around'' the circumnuclear dust and thereby diminishes the strong absorption efficiency of the flared dusty disk. 

When the winds are added, we find longer time lags with a more shallow dependence on viewing angle than for the baseline model. This is related to the fact that before escaping, the photons cross a second scattering region that adds more scattering events and increases the light travel path. The polarization position angle remains at 0$^{\circ}$ over all viewing angles for models with higher-density polar winds.
This behavior, linked to the optical depth of the polar outflows, is in agreement with the polarization state observed for six Seyfert 1 AGN by \citet{2011ApJ...738...90B}. Since \citet{2002MNRAS.335..773S}, it has been a well known fact that a small fraction of type 1s shows optical polarization spectra similar to those of Seyfert 2 galaxies in which polarized broad lines are detected. In the compendium of AGN presented in \citet{2014MNRAS.441..551M}, it was shown that these peculiar objects are associated with polarizations of a few percent and represent a rather small subclass of Seyfert
1s.\\
Similarly to our previous conclusion on the dependence of optical depth versus inclination, we note that a clumpy dust distribution is not very sensitive to inclination angles. Across type-1 and moderate type 2 viewing angles, the time lag distribution remains rather flat for both geometrical clumpy configurations; the time lag only increases for extreme type 2 lines of sight. This ambiguity is stronger for the wind with higher optical depth. In contrast to this, the time lag remains a discriminator between type 1 and type 2 inclination for the flared-disk geometry and is far higher with a uniform dust distribution.

\subsection{Adding a scattering ring to the baseline model}
\label{Ring}
We now include a third scattering region, an equatorial, ionized disk located between the source and the circumnuclear region. Compared to a model without 
equatorial electron disk (Fig.~\ref{fig:no_winds}), our new modeling shows that only the polarization properties of light and its time lags are affected (see Fig.~\ref{fig:baselineEQ}). An additional scattering component that does not absorb radiation has almost no effect on the flux level observed at various inclinations. However, the electron disk stabilizes the degree of 
polarization observed in type 1 viewing angles, independently of the circumnuclear dust structure. The disk also sets the polarization of radiation to 90$^\circ$ at type 1 orientations, as has been
observed \citep{2004MNRAS.350..140S}. More importantly, the proximity of this region to the central source means that the observed time lags are shorter for type 1 inclinations. A fraction 
of the input radiation, depending on the solid angle covered by the opening angle of the electron ring, is directly scattered toward the observer, inducing a shorter time lag. For higher inclinations, 
radiation passes through the optically thin medium and scatters inside the optically thick dust structure, so that the effect of the equatorial ring becomes almost undetectable. 

When we include polar outflows in addition to this scattering ring, see Fig.~\ref{fig:windsEQ}, we reach similar conclusions. The addition of an equatorial electron ring remains marginal 
in terms of total fluxes, but strongly helps the polarization degree and angle to stabilize at typical values seen for local AGN (see \citealt{2014MNRAS.441..551M}). The degree of polarization remains lower 
than 1\% for type 1 inclinations and sharply rises for type 2 orientations. The polarization angle rotates from parallel to perpendicular as soon as the observer's line of sight is obscured by 
the circumnuclear dusty region. The effect of the disk on the time lags at type 1 views is even stronger: the time lags are 5 -- 10 times shorter, reflecting the presence of the 
ring. The effect of different wind optical depths on time lags is the same as what we described in Sect.~\ref{Polar_outflows}.

\section{Summary and conclusion}
\label{sec:discuss}

In this initial study, we extended polarization modeling of AGN into the timing domain. Investigating the time lag of the polarized emission as a function of viewing angle, we showed that owing to multiple scattering effects, the polarized signal in type
2 objects lags farther behind the total intensity than for type-1 sources. Furthermore, the geometrical shape of the dust distribution such as the shape of the inner surfaces and also their optical depth has an effect on the resulting time delay at type 2 lines of sight, and much less for type 1. A significant increase in time delay is recorded when the dust distribution is clumpy.

When adding polar winds to the model, scattered radiation further increases the polarization time lag. The presence of a scattering equatorial electron disk naturally reduces the observed time lags at type 1 inclinations since a fraction of the photon flux is reprocessed onto the electrons and escapes along the
polar directions. 

The overall picture thus remains complex and certainly still includes degeneracy. From the modeling we present here, it is not yet straightforward, for instance, to disthinguish between the density structure of polar outflows and the level of clumpiness in the torus. Nonetheless, treating polarization as a function of time offers another independent way to constrain the model parameters. For this, it is important to combine the polarization time lag, polarization degree, and position angle and compare these observables to the type of simulation results that we present here.

This study is inspired by the discovery of polarization reverberation mapping in NGC~4151 \citep{2012ApJ...749..148G}. In this paper we also refer to the initial modeling of polarization time lags with {\sc stokes}. The various uniform-density tori that were tested at that time revealed that the polarization time lag at type 1 viewing angles is only slightly longer than the light-crossing time of the inner torus radius. In the following the torus was ruled out as a reverberation source because the observed time lags were shorter by a factor of five. They instead corresponded to the light-crossing time of the BLR. Polarization reverberation contains important seismological information about AGN, but as for polarization in general, careful modeling is required to correctly interpret it. Next, more central structures like the BLR and the outer parts of the accretion disk need to be included to recover shorter reverberation times. This work is currently underway and will be presented in a forthcoming publication, including dependencies on the density of the circumnuclear region and the cross correlation between polarized and unpolarized flux.

In the meantime, we would like to advocate systematic polarization monitoring campaigns of AGN such as those by \citet{2014MNRAS.440..519A,2015MNRAS.448.2879A}. Although these are long-term projects, they have a high potential to reveal more details of the emission and scattering geometry in AGN.

\begin{acknowledgements}
We would like to thank the referee for useful comments and suggestions that improved our paper. We would also like to acknowledge Jean-Marie Hameury for his additional remarks. 
{This research was supported by the CONICYT BECAS Chile grant no. 72150573 and the French Agence Nationale de la Recherche through the POLIOPTIX grant ANR-11-JS56-013-01 and the French Programme Nationale des Hautes Energies (PNHE). We are grateful to the French Government and the French Embassy in Serbia and the Ministry of Education and Science (Republic of Serbia) through the project Astrophysical Spectroscopy of Extragalactic Objects (176001) }
\end{acknowledgements}

\bibliography{bibliography}
\bibliographystyle{aa} % style aa.bst
%\bibliography{bibliofile}      

%\input{bibliography_manual.tex}

\end{document}